\documentclass[aps,prl,reprint,amsmath,amssymb]{revtex4-1}

\usepackage{graphicx}
\usepackage{dcolumn}
\usepackage{bm}
\usepackage{color}

\newcommand{\black}[1]{\color{black}#1}
\newcommand{\exciting}{{\usefont{T1}{lmtt}{b}{n}exciting}}
\begin{document}

\title{Work-function modification of PEG(thiol) adsorbed on the Au(111) surface: \\ A first-principles study}
\author{Jongmin Kim}
\author{Andris Gulans}
\author{Claudia Draxl}
\altaffiliation[]{claudia.draxl@physik.hu-berlin.de}
\affiliation{Institut f{\"u}r Physik and IRIS Adlershof, Humboldt-Universit{\"a}t zu Berlin, 12489 Berlin, Germany}

\date{\today}

\begin{abstract}
The possibility of modifying the work function of electrodes is important for optimizing the energy barriers for charge-injection (extraction) at the interface to an organic material. In this study, we perform density-functional-theory calculations to investigate the impact of dithiol-terminated polyethylene glycol (PEG(thiol)) based self-assembled monolayers (SAMs) with different numbers of PEG repeat units on the work function of the Au(111) surface. We find that a monolayer of PEG(thiol) decreases the work function of the Au(111) surface, where the magnitude of this reduction strongly depends on the length of the PEG backbone. The main contribution arises from the dipole due to the adsorption-induced charge rearrangement at the interface. Our work reveals a pronounced odd-even effect that can be traced back to the dipole moment of the PEG(thiol) layer. 
\end{abstract}

\maketitle

\section{Introduction}
Organic electronics devices, such as organic light emitting diodes (OLEDs), organic photovoltaics (OPVs), and organic field-effect transistors (OFETs), are heavily affected by physical phenomena at the interface between the electrode and the organic material \cite{Parker1994, Ishii1999}. Often, the design of such devices faces fundamental challenges due to poor charge-injection (extraction) \cite{Cheng2009, Liu2015, Kim2014, Choi2016}. Reducing the corresponding energy barrier, can considerably improve device performance, and a number of approaches has been reported. Among them are doping of the organic semiconductor \cite{Walzer2007} or the modification of the work function of the electrode \cite{Chuang2014}.

For optimal charge-injection, the work function of the anode needs to be closely aligned to the highest occupied molecular orbital (HOMO) of the organic semiconductor. Likewise, the work function of the cathode requires to be matched with the lowest unoccupied molecular orbital (LUMO) level. In other words, anode and cathode should be made of materials with high and low work functions, respectively \cite{Heimel2008}. Metals with a low work function, which are typically used for the cathode such as Ca, Mg, and Al \cite{Bloom2003, Matz2013, Zhou2012}, are, however, immensely reactive and oxidize due to moisture or oxygen which results in an instability of devices \cite{Zhou2012,Jorgensen2008}. Chemically inert metals, such as Au and Ag, have a high work function, and thus large charge-injection (extraction) energy barriers \cite{Zhou2012}, and are therefore difficult to be used as the cathode. This problem can be addressed by introducing an interlayer between the cathode and the organic semiconductor to adjust the work function of the electrode \cite{Braun2009, Khan2014, Ratcliff2011, Koch2007}. Materials that have been employed as interlayers are mainly polymers, metal oxides, inorganic salts, and self-assembled monolayers (SAMs), which modulate the electrode's work function by inducing a dipole at the interface \cite{Zhou2012, Kano2009, Chen2008, Cheng2009, Kim2014, deBoer2005}.

Recent experimental studies have shown that polyethylene glycol (PEG) utilized as the interlayer improves the performance of OPVs, OLEDs, and OFETs \cite{Deckman2015, Shamieh2016, Vinokur2017, Nouzman2017, Sarkar2020}. In addition, experimental works \cite{Deckman2015, Shamieh2016, Vinokur2017, Nouzman2017, Sarkar2020, Shamieh2018} have demonstrated that PEG-based additives blended with the organic semiconductor are able to migrate to the interface, forming a self-generated interlayer by the interaction between the head groups of the additives and the electrodes. Although this technique is applicable to fabricating devices, it is difficult to accurately measure the energy-level alignment or the structural conformation because the interlayer is formed at a buried interface \cite{Shamieh2018}. Alternatively, one could consider a SAM of PEG molecules as the interlayer. On the one hand, interfaces between a SAM and an electrode would be easier to analyze. On the other hand, the electronic properties of the interface could be chemically tuned, thus exploiting a common advantage of SAMs. SAMs based on alkanethiols and phenylthiols have been intensively investigated in view of modifying the properties of Au electrodes \cite{deBoer2005, Love2005, Tao2007, Lee2015, Heimel2008, Heimel2010}. However, studies on an adsorbed SAM of PEG on the electrode material and its impact on the modulation of the work function are still lacking. 

In the present work, we perform first-principles calculations to model dithiol-terminated PEG (PEG(thiol)) deposited on the Au(111) surface. 
We show how such a SAM decreases the metal's work function. In addition, we also show that the work function modification is sensitive to the length of the PEG backbone.
We demonstrate that a pronounced odd-even effect originates from the relative orientation of the molecular PEG(thiol) dipole with respect to the surface normal.
{
}

\begin{figure}[hb]
\begin{center}
\includegraphics[width=1.0 \linewidth]{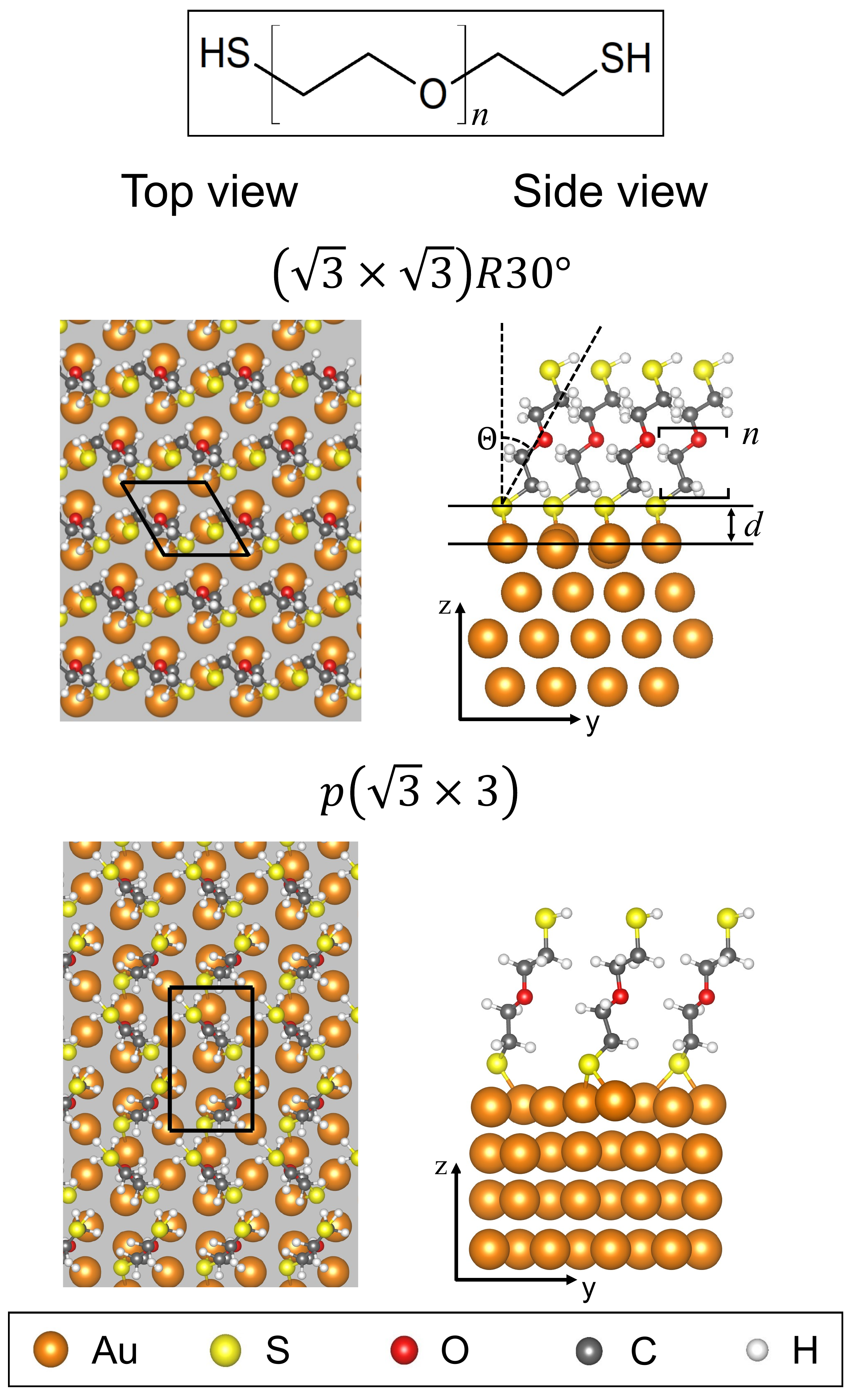}
\end{center}
\caption{Top: Chemical structure of the PEG(thiol) molecule. $\textit{n}$ = 1, 2, 3, and 4 is the number of repeat units of the PEG backbone. {\black Top and side view of a monolayer of PEG(thiol) molecules adsorbed on the Au(111) surface in $(\sqrt3 \times \sqrt3)R30^\circ$ (middle) and $p$($\sqrt{3}$ $\times$ 3) (bottom) unit cells. The surface unit cells are indicated by the black lines. $\theta$ is the tilting angle between the PEG backbone and the surface normal. $\textit{d}$ is the height from the S atom to the average position of the surface atoms.}
}
\label{fgr:figure1}
\end{figure}

\section{Computational details}
We model the PEG(thiol) molecules adsorbed on Au(111), termed PEG(thiol)@Au(111), for different numbers of repeat units of the PEG backbone (1 to 4). 
The interface structure is shown in Fig.~\ref{fgr:figure1}. 
The adsorption geometry generally depends on the molecular packing density on the surface. We consider the high-coverage limit, where the PEG(thiol) molecules form well-ordered van-der-Waals bonded SAMs~\cite{Love2005}. Scanning tunneling microscopy (STM) and grazing incidence X-ray diffraction (GIXD) experiments of thiolate adsorbed on Au(111) show coexistance of a $(\sqrt3 \times \sqrt3)R30^\circ$ and a c(4 $\times$ 2) overlayer structure \cite{Vericat2001, Vericat2005}, 
where the ratio of their concentration depends on molecular length and temperature \cite{Vericat2005}. In particular, for longer molecules the $(\sqrt3 \times \sqrt3)R30^\circ$ is preferred. We thus adopt the $(\sqrt3 \times \sqrt3)R30^\circ$ surface unit cell.

{\black For the sake of comparison, we also consider a $p$($\sqrt{3}$ $\times$ 3) unit cell that contains two PEG molecules forming a herringbone pattern, which has been observed in a previous study of thioaromatic monolayers on Au(111) \cite{Azzam2003}.
It has the same coverage as the $(\sqrt3 \times \sqrt3)R30^\circ$ overlayer.
Its structure is displayed in the bottom panel of Fig.~\ref{fgr:figure1}.
}
In all considered systems, the metal substrate consists of four atomic layers.
The positions of the Au aconcentration toms in the two top layers and the adsorbed PEG(thiol) are relaxed until the maximum force on each atom is smaller than 0.055 eV/$\text{\AA}$. The bottom two layers are fixed to the positions corresponding to the bulk structure with a lattice constant of 4.192~$\text{\AA}$. The vacuum spacing along the vertical direction is at least 14~$\text{\AA}$ to avoid spurious interactions between the periodic images.

We perform density-functional theory (DFT) calculations using the all-electron full-potential code \exciting\ \cite{Gulans2014}. It employs the linearized augmented plane-wave plus local orbitals (LAPW+lo) basis. The muffin-tin radii are set to $R_{MT}^{Au}=$~2.1 bohr, $R_{MT}^{S}=$~1.6 bohr, $R_{MT}^{O}=$~1.1 bohr, $R_{MT}^{C}=$~1.1 bohr, and $R_{MT}^{H}=$~0.7 bohr, respectively. The planewave cutoff $G_{max}$ in the interstitial region is set to 4.29 bohr$^{-1}$ which corresponds to a value of $R_{MT}G_{max}=$~3 for hydrogen (having the smallest muffin-tin sphere) and $R_{MT}G_{max}=$~9 for Au. {\black The Brillouin zone (BZ) is sampled on a 5 $\times$ 5 $\times$ 1 and a 5 $\times$ 3 $\times$ 1 \textbf{\textit{k}}-point mesh for the $(\sqrt3 \times \sqrt3)R30^\circ$ and the $p$($\sqrt{3}$ $\times$ 3) surface unit cells, respectively.}
The energy convergence criterion for the self-consistency is set to $10^{-6}$ Hartree.

Exchange and correlation effects are described by means of the generalized gradient approximation (GGA) as parametrized by Perdew-Burke-Ernzerhof (PBE) \cite{Perdew1996}, with corrections for the van der Waals interactions on top. These long-range correlation effects, which have a strong impact on the adsorption structures in numerous cases, are considered by two different schemes, namely by DFT-D2 \cite{Grimme2006} and by the many-body disperison method (MBD@rsSCS) \cite{Ambrosetti2014,Bucko2016}. Both methods are implemented in the \exciting\ code.

DFT-D2 employs a pairwise additive model to compute the long-range dispersion energy which is expressed as
\begin{equation}
E_{disp}=-\frac{1}{2} s_{6} \sum_{AB} \frac{C_{6,AB}}{(r_{AB})^6} f_d(r_{AB}).
\end{equation}
Here $s_{6}$ is a global scaling factor for the dipole-dipole dispersion coefficients $C_{6,{AB}}$, and $r_{AB}$ is the interatomic distance between atoms A and B. The $C_{6,{AB}}$ coefficients are calculated as a geometrical mean of fixed empirical coefficients for the atoms, and $f_d$ is a damping function.

In contrast, MBD@rsSCS goes beyond pairwise interactions, and it also includes screening effects.
To calculate the dispersion energy, one first evaluates the atomic polarizability, $\alpha^{\mathrm{TS}}$, from the polarizability of the free atoms, scaled by their volumes that are estimated via Hirshfeld partitioning ($\alpha^{\mathrm{TS}}=\alpha^{free} \ \! {V^{eff}}/{V^{free}}$) as suggested by Tkatchenko and Scheffler (TS) \cite{Tkatchenko2009}. Then the short-range part of the dipole-dipole interaction is considered to compute the screened polarizability, $\bm{\tilde{\alpha}}^{\mathrm{SCS}}$, using $\alpha^{\mathrm{TS}}(\omega)$. To this extent, the short-range screening equation is solved self-consistently:
\begin{equation}
\label{alphascs}
\bm{\tilde{\alpha}}^{\mathrm{SCS}}(\omega)=\alpha^{\mathrm{TS}}(\omega)\left(\textbf{1}- \textbf{T}_{SR}(\omega) \bm{\tilde{\alpha}}^{\mathrm{SCS}}(\omega)\right).
\end{equation}
Here $\textbf{T}_{SR}$ is the short-range dipole-dipole interaction tensor. The long-range counterpart, $\textbf{T}_{LR}$, is computed as 
\begin{equation}
\label{TLR}
\textbf{T}_{LR,AB}^{ab}(\textbf{\textit{k}})=\sum \limits_{\textbf{\textit{L}}} f(\tilde S_{\mathrm{vdW},AB},r_{AB,\textbf{\textit{L}}})\textbf{T}_{AB,\textbf{\textit{L}}}^{ab}\cdot\mathrm{e}^{-\mathrm{i}\textbf{\textit{k}}\cdot\textbf{\textit{L}}},
\end{equation}
where
\begin{equation}
\label{Tabij}
\textbf{T}_{AB}^{ab}=\frac{3r_{AB}^{a}r_{AB}^{b}-\delta_{ab}r_{AB}^{2}}{r_{AB}^{5}}
\end{equation}
is the second-order interaction tensor and $f$ is a Fermi-type damping function. $\textbf{\textit{L}}$ is a translation vector, $r_{AB}$ is the distance between atoms $A$ and $B$, and $a$, $b$ indicate Cartesian components. $\tilde S_{\mathrm{vdW}}$ is the sum of the van der Waals radii of the screened atoms, $\tilde S_{\mathrm{vdW}}=\beta (\tilde R_{\mathrm{vdW},A}+\tilde R_{\mathrm{vdW},B}$) scaled by a parameter $\beta$.
Finally, the dispersion energy becomes \cite{Bucko2016}
\begin{equation}
\small
E_{disp}=-\sum_{\textbf{\textit{k}}}w_{\textbf{\textit{k}}}\int_0^{\infty} \frac{d\omega}{2\pi} \mathrm{Tr}\{ln(\textbf{1}-\textbf{A}_{LR}(\omega) \textbf{T}_{LR}(\textbf{\textit{k}}))\},
\end{equation}
where $w_{\textbf{\textit{k}}}$ represents the weight of a \textbf{\textit{k}}-point, and $\textbf{A}_{LR}$ is 1/3 of the trace of $\bm{\tilde{\alpha}}^{\mathrm{SCS}}$. For structure optimizations involving the long-range correlation in the MBD@rsSCS method, interatomic forces are computed from the dispersion-energy gradient.

\section{Results and discussion}

\subsection{Adsorption geometry}
The main structural features of the relaxed PEG(thiol)@Au(111) system are reported in Table~\ref{tbl:table1}. The molecules are chemically bound to the surface through their S atom of the head group, situated at the bridge site with a slight shift toward the hollow site as can be seen in top views depicted in Fig.~\ref{fgr:figure1}. This adsorption configuration is similar to what was found by various theoretical studies of thiolate adsorbed on Au(111) \cite{Nara2004, Tonigold2013, Tonigold2015}. In our MBD@rsSCS calculations, the PEG(thiol) molecules are tilted from the surface normal by 29.5 -- 31.9$^\circ$ depending on the number of repeat units for the $(\sqrt3 \times \sqrt3)R30^\circ$ pattern. 
{\black Likewise, the two molecules in the $p$($\sqrt{3}$ $\times$ 3) pattern are adsorbed with tilting angles of 33.6$^\circ$ and 33.3$^\circ$, that is, 33.5$^\circ$ on average.}
Similar tilting angles for n-alkanethiols on the Au surface are reported in Refs. 41 and 42. The calculated adsorption heights, \textit{d}, between the S atom and the average position of Au surface atoms are 1.93 -- 1.95~$\text{\AA}$, depending on the molecular length. In contrast, the tilting angles and adsorption heights, are not sensitive to the latter.
{\black 
The $p$($\sqrt{3}$ $\times$ 3) exhibits adsorption heights of 1.98~$\text{\AA}$ and 1.94~$\text{\AA}$ of the two molecules, average adsorption height of 1.96~$\text{\AA}$.
These structural parameters are similar in both patterns.
On the other hand, the adsorbed molecules in the herringbone pattern
are less bent compared to those in the $(\sqrt3 \times \sqrt3)R30^\circ$, thus, they are in an upright conformation.}


\begin{table}[t]
\small
\caption{Adsorption geometry of PEG(thiol) molecules adsorbed on the Au(111) surface using MBD@rsSCS and DFT-D2 for van der Waals corrections on top of PBE (see also Fig.~\ref{fgr:figure1}).}
\label{tbl:table1}
\begin{tabular*}{0.45\textwidth}{@{\extracolsep{\fill}}ccccc}
\hline
\hline
\\[-1em]
Repeat units  &\multicolumn{2}{c}{MBD@rsSCS}   & \multicolumn{2}{c}{DFT-D2}\\
\cline{2-3}\cline{4-5}
\\[-1em]
$\textit{n}$         & $\theta$ [$^\circ$]   & $\textit{d}$ [$\text{\AA}$]   & $\theta$ [$^\circ$]   & $\textit{d} $[$\text{\AA}$] \\
\hline
\\[-1em]
&\multicolumn{4}{c}{$(\sqrt3 \times \sqrt3)R30^\circ$}\\
\cline{2-5}
\\[-1em]
 1     &  31.9    & 1.94    &  35.3   & 1.92  \\
 2     &  29.5    & 1.95    &  32.0   & 1.93  \\
 3     &  31.3    & 1.94    &  32.3   & 1.92  \\
 4     &  30.7    & 1.93    &  30.7   & 1.92  \\
\hline
\\[-1em]
&\multicolumn{4}{c}{{\black $p$($\sqrt{3}$ $\times$ 3)}}\\
\cline{2-5}
\\[-1em]
1     &  {\black 33.5}    & {\black 1.96}    &  -   & -\\
\hline
\hline
\end{tabular*}
\end{table}

\subsection{Work-function change}
\begin{figure}[b]
\begin{center}
\includegraphics[width=1.0 \linewidth]{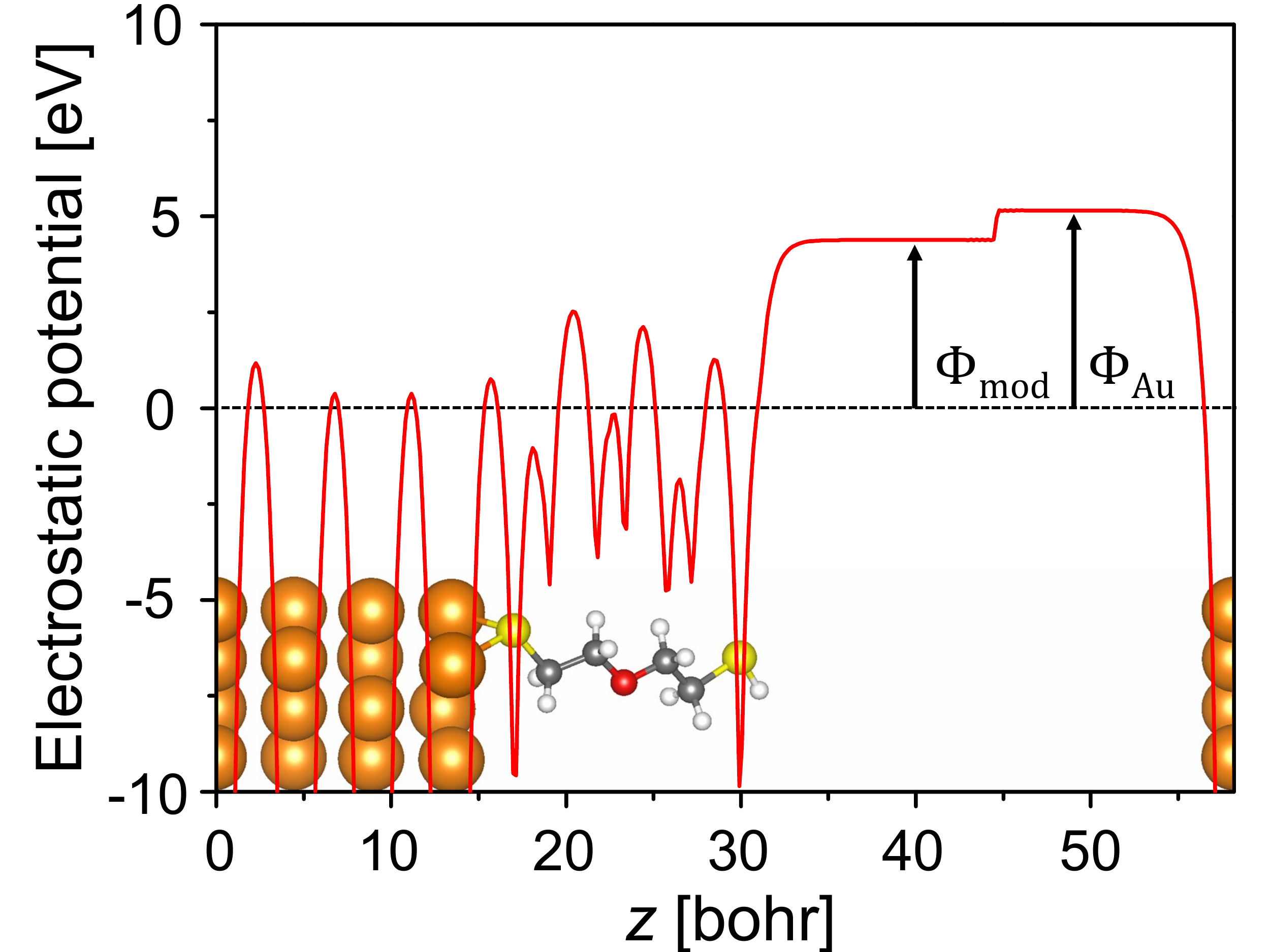}
\end{center}
\caption{Plane-averaged electrostatic potential of the hybrid system, consisting of a PEG(thiol) monolayer adsorbed on Au(111), obtained with the MBD@rsSCS approach. The Fermi level is set to zero.}
\label{fgr:figure2}
\end{figure}

In a next step, we study the influence of the adsorbed PEG(thiol) molecules on the work function of Au(111). 
Figure \ref{fgr:figure2} displays the calculated plane-averaged electrostatic potential of the investigated system obtained by MBD@rsSCS {\black for the $(\sqrt3 \times \sqrt3)R30^\circ$ case with $\textit{n}=1$} (1-PEG(thiol)@Au(111)). The work function is defined as the difference of the electrostatic potential energy at the vacuum level, $E_{vac}$, and the Fermi energy, $E_{f}$, i.e. $\Phi=E_{vac}-E_{f}$.

Note that the two sides of the slab have different values of $E_{vac}$. The side with the clean Au(111) surface reflects the work function of gold, $\Phi_{Au}$, for which our calculation yields a value of $\Phi_{Au}= $ 5.15~eV. It agrees well with 5.15~eV obtained from an UPS experiment \cite{Duhm2008} as well as previous DFT-PBE studies where values of 5.15~eV \cite{Miller2009} and 5.12~eV \cite{Patra2017} were reported. The other side of the slab has the work function, modified by the molecular adsorption, $\Phi_{mod}=$ 4.39~eV. Thus, the effect of the adsorbed SAM is given by the difference $\Delta\Phi$ = $\Phi_{mod}-\Phi_{Au}=$~-0.76~eV. This quantity is presented in Table~\ref{tbl:table2} as a function of $\textit{n}$. We observe that (i) the PEG(thiol) monolayer significantly decreases the work function of Au(111) in all cases and (ii) there are strong oscillations of $\Delta\Phi$ with respect to $\textit{n}$. An odd (even) number of repeat units results in the smallest (largest) work-function modification. This phenomenon is known as an odd-even effect \cite{Love2005,Tao2007}.

\begin{figure*}
\begin{center}
\includegraphics[width=1.0 \linewidth]{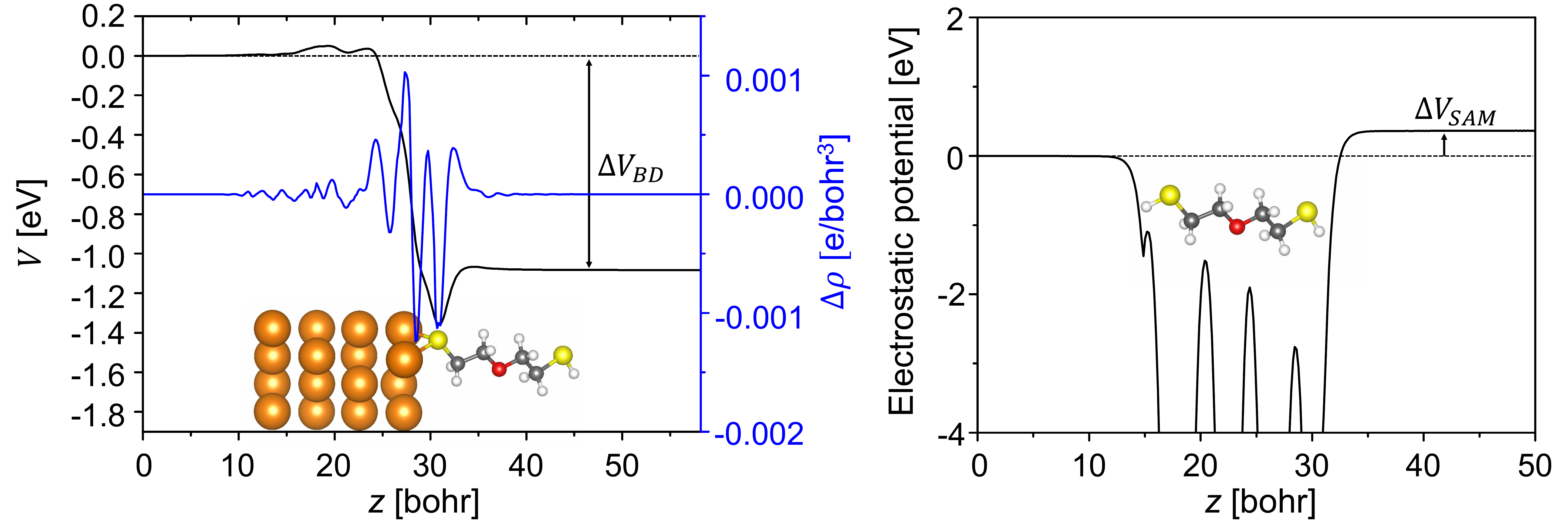}
\end{center}
\caption{Left: Plane-averaged charge rearrangement, $\Delta\rho$, (right axis) and corresponding change in potential energy (left axis), due to the bond dipole (BD) induced by adsorption of 1-PEG(thiol). A positive (negative) value of $\Delta\rho$ represents accumulation (depletion) of charge density. Right: Electrostatic potential of an isolated monolayer of 1-PEG(thiol). All results are obtained with the MBD@rsSCS approach.}
\label{fgr:figure3}
\end{figure*}

\begin{table*}
\caption{PEG(thiol)-induced change in work function, $\Delta\Phi$, and its main components for different numbers of repeat units of the PEG backbone, $\textit{n}$. All results obtained using MBD@rsSCS and DFT-D2, respectively, are given in eV.}
\label{tbl:table2}
\begin{tabular*}{0.9\textwidth}{@{\extracolsep{\fill}}cccccccccc}
\hline
\hline
Repeat units   &\multicolumn{4}{c}{MBD@rsSCS}     & \multicolumn{4}{c}{DFT-D2}\\
\cline{2-5}\cline{6-9}
\\[-1em]
$\textit{n}$       &$\Delta\Phi$   & $\Delta V_{BD}$     &$\Delta V_{SAM}$  &$\Delta V_{relax-Au}$   &$\Delta\Phi$   & $\Delta V_{BD}$     &$\Delta V_{SAM}$  &$\Delta V_{relax-Au}$   \\
 \hline
 \\[-1em]
&\multicolumn{8}{c}{$(\sqrt3 \times \sqrt3)R30^\circ$}\\
\cline{2-9}
\\[-1em]
 1       &  -0.76        & -1.08    & 0.36    &-0.04    &  -0.24        & -1.01    & 0.79    &-0.02\\
 2       &  -1.11        & -1.06    & 0.00    &-0.05    &  -0.82        & -1.00    & 0.22    &-0.04\\
 3       &  -0.39        & -1.12    & 0.78    &-0.05    &  -0.26        & -1.06    & 0.84    &-0.04\\
 4       &  -1.16        & -1.10    & 0.00    &-0.06    &  -0.95        & -1.02    & 0.08    &-0.01\\
\hline
\\[-1em]
&\multicolumn{8}{c}{{\black $p$($\sqrt{3}$ $\times$ 3)}}\\
\cline{2-9}
\\[-1em]
1      &  {\black -0.37}        & {\black -0.96}    & {\black 0.64}    & {\black -0.05}    &  -        & -    & -    &-\\
\hline
\hline
\end{tabular*}
\end{table*}

In order to analyze the nature of the work-function change, we perform an analysis following Refs. 46 -- 48 and decompose $\Delta\Phi$, as follows:
\begin{equation}
\label{eq:DeltaPhi}
\Delta\Phi=\Delta V_{BD}+\Delta V_{SAM}+\Delta V_{relax-Au}.
\end{equation}
$\Delta V_{BD}$ is the contribution due to the charge rearrangement caused by the formation of new chemical bonds which leads to a dipole between PEG(thiol) and Au(111). $\Delta V_{SAM}$ is the shift of the electrostatic potential created by the intrinsic dipole moment of the PEG(thiol) layer relative to the surface normal. $\Delta V_{relax-Au}$ indicates the work function change of the isolated Au surface due to the surface relaxation caused by the adsorption of PEG(thiol). We define $\Delta V_{SAM}$ as the difference of the electrostatic potential energy between the two sides of the PEG(thiol) molecules. Likewise, the difference of the potential energy between the two sides of the isolated Au(111) slab is termed $\Delta V_{relax-Au}$.

To obtain $\Delta V_{BD}$, we consider the change of the charge density caused by the chemical bonding. The latter induces a change in the electrostatic potential, $\Delta V$, which satisfies the Poisson equation. In practice, it is sufficient to consider a plane-averaged density change, $\Delta\rho$, to solve
\begin{equation}
\frac{d^2\Delta V}{dz^2}=-4\pi\Delta\rho.
\end{equation}
by numerical integration. The difference of $\Delta V$ between two sides of the slab corresponds to $\Delta V_{BD}$ (see Fig.~\ref{fgr:figure3}).

In PEG(thiol)@Au(111), when PEG(thiol) molecules are attached to the Au surface, the S--H bond of thiol is replaced by an S--Au bond, followed by release of H$_2$ \cite{Love2005, Ulman1996}. Therefore, $\Delta\rho$ is defined as
\begin{equation}
\Delta\rho= \rho_{tot}-(\rho_{SAM}+\rho_{surf}-\rho_{H}),
\end{equation}
where $\rho_{tot}$, $\rho_{SAM}$, $\rho_{surf}$, and $\rho_{H}$ correspond to the plane-averaged charge densities of the total PEG(thiol)@Au(111) system, the free-standing PEG(thiol) monolayer, the isolated Au(111) surface from the total system, and the isolated layer of H atoms, respectively.

Note that Eq.~\ref{eq:DeltaPhi} is exact with above definitions of $\Delta\Phi$, $\Delta V_{BD}$, $\Delta V_{SAM}$, and $\Delta V_{relax-Au}$. If all four terms are expressed in terms of dipole moments that can be calculated from the corresponding nuclear charges and self-consistent densities, it becomes apparent that Eq.~\ref{eq:DeltaPhi} holds strictly as long as the isolated layer of H atoms does not have a dipole moment along the $z$-direction, which is indeed the case due to symmetry.

\begin{figure}
\begin{center}
\includegraphics[scale=0.32]{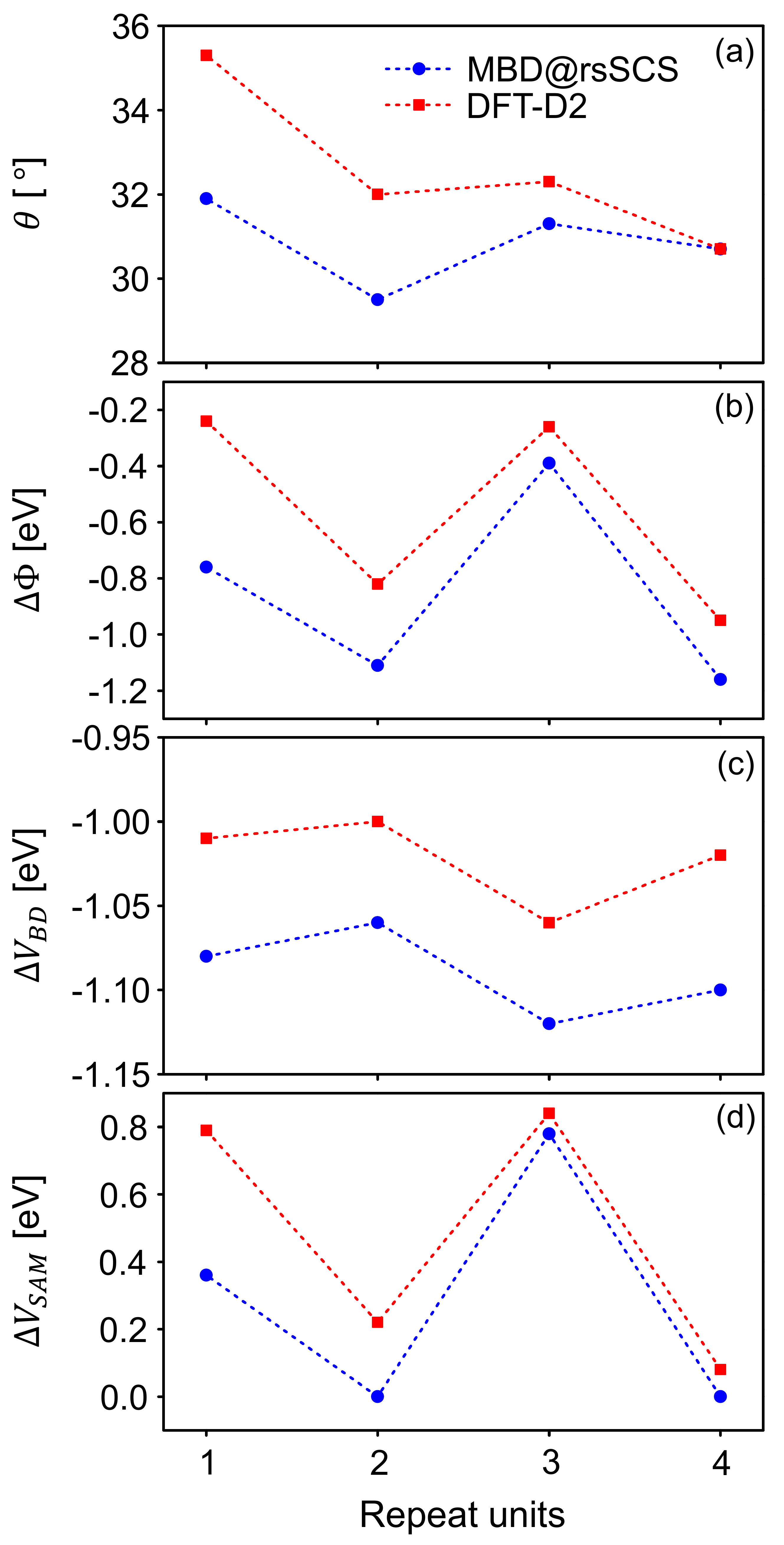}
\end{center}
\caption{Structural and electronic properties of a monolayer of PEG(thiol) molecules adsorbed on Au(111) as a function of repeat units, $\textit{n}$, obtained by DFT-D2 (red) and MBD@rsSCS (blue): (a) tilting angle, $\theta$, (b) modification in work function, $\Delta\Phi$, (c) change in potential-energy at the interface, $\Delta V_{BD}$, and (d) potential-energy shift along the PEG(thiol) molecules.}
\label{fgr:figure4}
\end{figure}

Figure~\ref{fgr:figure3} provides the results of $\Delta V_{BD}$ and $\Delta V_{SAM}$ obtained with MBD@rsSCS for the case of 1-PEG(thiol)@Au(111) (shown in Fig.~\ref{fgr:figure2}). The left panel shows the plane-averaged $\Delta\rho$ and $\Delta V_{BD}$ along the vertical direction. Significant oscillations of the former are found at the interface. In other words, the charge rearrangement is mainly confined to the interface region, in particular, near the S atom and the topmost Au layer. The charge density is increased in the topmost Au layer and depleted right above it. For 1-PEG(thiol)@Au(111), the individual contributions to $\Delta\Phi$ amount to $\Delta V_{BD}=-1.08$~eV, $\Delta V_{SAM}=0.36$~eV, and $\Delta V_{relax-Au}=-0.04$~eV. 
The dominant contribution in $\Delta\Phi$ is $\Delta V_{BD}$, whereas $\Delta V_{relax-Au}$ is tiny compared to $\Delta V_{BD}$ and $\Delta V_{SAM}$. 
The formation of the bond dipole is accompanied by charge transfer. To quantify its amount, we integrate $\Delta \rho$ along the vertical direction \cite{Heimel2010, Wang2019, Romaner2009}, obtaining 0.11 electrons that are transferred from the S atom to the Au surface.

Let us now examine the dependence of the three contributions on the molecular length, $\textit{n}$, and the origin of the observed odd-even effect in $\Delta\Phi$. The corresponding values of $\Delta\Phi$, $\Delta V_{BD}$, $\Delta V_{SAM}$, and $\Delta V_{relax-Au}$ are listed in Table~\ref{tbl:table2}. $\Delta V_{BD}$ decreases the work function of the surface in all four cases. However, in contrast to $\Delta\Phi$, $\Delta V_{BD}$ has similar values irrespective of $\textit{n}$, which therefore has basically no influence on the odd-even effect. $\Delta V_{relax-Au}$ is also insensitive to $\textit{n}$, and its magnitude is negligible compared to $\Delta V_{BD}$ and $\Delta V_{SAM}$. This indicates that the surface relaxation upon adsorption does not play a role in the odd-even effect. Unlike $\Delta V_{BD}$ and $\Delta V_{relax-Au}$, a pronounced odd-even effect is revealed for $\Delta V_{SAM}$. It should be noted that values of $\Delta V_{SAM}$ positive potential shifts are obtained in case of odd $\textit{n}$ while for even $\textit{n}$ they are basically zero. The latter is caused by PEG(thiol) with even $n$ having an inversion center that cancels out the effect of the molecular dipole moment. Therefore, we attribute the odd-even effect to the PEG(thiol) dipole moment perpendicular to the surface, while the bond dipole contribution due to the adsorption-induced charge redistribution as well as the surface-relaxation contribution due to adsorption are relatively small.

The trend in the work-function modification for n-alkanethiols and CF$_3$-terminated n-alkanethiols on Au was also observed in previous studies, and it has been shown to originate from differences in the terminal dipole orientations of the molecular monolayer \cite{Tao2007, Lee2015}. Interestingly, the differences of $\Delta\Phi$ between odd and even chains of n-alkanethiols and CF$_3$-terminated n-alkanethiols are noticeably lower than those of the PEG(thiol)s \cite{Lee2015}. For example, the maximum amount of such energy differences is approximately 0.30~eV for the former and 0.77~eV for the latter.

Since van der Waals (vdW) forces play a crucial role at organic/metal interfaces \cite{Sony2007,Romaner2009,Tkatchenko2010} as well as in the intermolecular interactions \cite{Nabok2008}, we finally investigate how sensitive the work-function modification is to the choice of the vdW correction. To this extent, we perform additional calculations by means of the semiempirical DFT-D2 functional \cite{Grimme2006}, which employs a pairwise-additive model account for the long-range dispersion. The results on the adsorption geometry and the work function modification are given in Fig.~\ref{fgr:figure4} and Table~\ref{tbl:table2}.

The tilting angles, $\theta$, calculated with the DFT-D2 are higher than those obtained with the MBD@rsSCS.
The difference is getting smaller with longer molecular length, and the two approaches give the same answer for $\textit{n}=4$. Since $\theta$ obtained from MBD@rsSCS is similar for all $n$ unlike from DFT-D2, and since shorter (longer) molecules exhibit smaller (larger) dielectric constant/screening, we conclude that screening effects (only considered in MBD@rsSCS) have a minor effect on $\theta$. The adsorption heights, $\textit{d}$, obtained from DFT-D2 and MBD@rsSCS are 1.92 -- 1.93~$\text{\AA}$ and 1.93 -- 1.95~$\text{\AA}$, respectively. These values clearly reflect that there are no noticeable differences between the two approaches.

$\Delta\Phi$ as computed with both approaches is shown at Fig. 4(b), exhibiting a qualitatively similar trend. Both types of calculations show a decrease in the work function for all four $n$, and a pronounced odd-even effect is observed. However, in the MBD@rsSCS case, a greater reduction of the work function is obtained when the PEG(thiol) molecules approach the Au(111) surface compared to the DFT-D2 method. In the MBD@rsSCS calculations, the difference of $\Delta\Phi$ between $\textit{n}=2$ and $\textit{n}=4$ is only 0.05~eV, whereas a much greater difference of 0.37~eV is observed between odd values of $\textit{n}$. 
For $\textit{n}=1$, the molecule is more twisted and bent than the others. Since the difference between odd molecules is only 0.02~eV in case of DFT-D2, we argue that the weaker screening of the short molecule plays a dominant role in the MBD@rsSCS results.

Figs. 4(c) and 4(d) show $\Delta V_{BD}$ and $\Delta V_{SAM}$. $\Delta V_{BD}$ obtained from both methods is found to be negative but the values are slightly different. In contrast, $\Delta V_{SAM}$ associated with the dipole moment of the PEG(thiol) monolayer is sensitive to the type of long-range corrections. The difference between the results from MBD@rsSCS and DFT-D2 reflects variations in the structure of the adsorbed PEG(thiol) molecules, including the tilting angle as mentioned before. For example, the values of $\Delta V_{SAM}$ obtained with DFT-D2 are non zero for even $n$, unlike those obtained with MBD@rsSCS. Since the molecules relaxed by DFT-D2 have a curved, banana-like shape, they exhibit a finite dipole moment in absence of inversion symmetry. This leads to the $\Delta V_{SAM}$ potential shift. Overall, despite many differences between the results of the two methods, we note that the main message of our paper does not change if the DFT-D2 is used instead of the MBD@rsSCS.

{\black 

Now we analyze the results of the work-function modification for the $p$($\sqrt{3}$ $\times$ 3) pattern. 
The work-function change and the corresponding contributions are reported in Table~\ref{tbl:table2}. 
Comparing only the results from the MBD@rsSCS calculations, the obtained $\Delta \Phi=-0.37$~eV is about half of that obtained for the $(\sqrt3 \times \sqrt3)R30^\circ$ pattern with $n=$~1.
This difference mainly originates from molecules in the herringbone pattern that are less curved unlike in the $(\sqrt3 \times \sqrt3)R30^\circ$ case. 
On the other hand, $\Delta \Phi$ is nearly identical to the work-function change in the $(\sqrt3 \times \sqrt3)R30^\circ$ system for $n=$~3. 
Analyzing the contributions to $\Delta \Phi$, we find that both, $\Delta V_{BD}$ and $\Delta V_{SAM}$, are slightly smaller in their magnitude in the herringbone pattern compared to the $(\sqrt3 \times \sqrt3)R30^\circ$ case.
Nevertheless, the differences in the individual terms, $\Delta V_{BD}$ and $\Delta V_{SAM}$, are canceled out resulting in the very similar total changes in the work-function. 

The MBD@rsSCS and DFT-D2 calculations performed for the $(\sqrt3 \times \sqrt3)R30^\circ$ pattern show that, as long as the adsorbed molecules are in an upright configuration and remain mostly undistorted, $\Delta \Phi$ is consistent for the values of $n$ with the same parity.
Therefore, we anticipate that PEG(thiol) in the herringbone pattern follows the same trend, and, moreover, both patterns yield similar work-function changes for $n>1$.
As discussed above, the $(\sqrt3 \times \sqrt3)R30^\circ$ and the $p$($\sqrt{3}$ $\times$ 3) structures can coexist on the Au surface forming a mixed disordered phase. 
Based on our results, we expect no noteworthy difference in the work-function compared to the calculated values for molecules with at least two repeat units. 

}

\section{Conclusions}
In summary, we have carried out first-principles calculations of the work-function modification induced by the adsorption of a PEG(thiol) SAM on the Au(111) surface, considering different numbers of PEG backbone repeat units. We have found that the PEG(thiol) molecules are adsorbed with an average tilting angle of the PEG backbone of $\sim$ 30$^\circ$ with respect to the surface normal. Importantly, the work function of Au(111) is always reduced, regardless of the molecular length, and the obtained reduction improves the charge-injection at the interface. We observe a pronounced odd-even effect in the work-function shift with the number of repeat units. An even number of repeat units reduces the work function of Au(111) more than an odd one. This effect stems mainly from a dipole moment of the PEG(thiol) molecules, and it should be considered when designing interfaces for applications in molecular electronics.
{\black 
We have compared two coverage patterns, $(\sqrt3 \times \sqrt3)R30^\circ$ and $p$($\sqrt{3}$ $\times$ 3), and conclude that they yield the same change of the work-function for a number of repeat units larger than one.
}


\acknowledgments
J.K. appreciates funding from the European Community's Horizon 2020 research and innovation program under Marie Sk{\l}odowska-Curie grant agreement No. 675867. C.D. and A.G. thank the Deutsche Forschungsgemeinschaft (DFG) for supporting Projektnummer 182087777 - SFB 951. C.D. acknowledges the inspiring scientific atmosphere and financial contributions from the Institute for Pure and Applied Mathematics (IPAM) at UCLA during the 2016 fall program on “Understanding Many-Particle Systems with Machine Learning”. We thank Basel Shamieh, Tanmoy Sarkar, and Gitti L. Frey for fruitful discussions. Input and output data can be downloaded from the NOMAD Repository, http://dx.doi.org/10.17172/NOMAD/2019.11.27-2.


\begin{thebibliography}{}
\bibitem{Parker1994} I. D. Parker, Appl. Phys. $\bf{75}$, 1656 (1994).
\bibitem{Ishii1999} H. Ishii, K. Sugiyama, E. Ito, and K. Seki, Adv. Mater. $\bf{11}$, 605 (1999).
\bibitem{Cheng2009} X. Cheng, Y.-Y. Noh, J. Wang, M. Tello, J. Frisch, R.-P. Blum, A. Vollmer, J. P. Rabe, N. Koch, and H. Sirringhaus, Adv. Funct. Mater. $\bf{19}$, 2407 (2009).
\bibitem{Liu2015} C. Liu, Y. Xu, and Y.-Y. Noh, Materials Today $\bf{18}$, 79 (2015).
\bibitem{Kim2014} J. Kim, Y. S. Rim, Y. Liu, A. C. Serino, J. C. Thomas, H. Chen, Y. Yang, and P. S. Weiss, Nano Lett. $\bf{14}$, 2946 (2014).
\bibitem{Choi2016} S. Choi, C. Fuentes-Hernandez, C.-Y. Wang, T. M. Khan, F. A. Larrain, Y. Zhang, S. Barlow, S. R. Marder, and B. Kippelen, ACS Appl. Mater. Interfaces $\bf{8}$, 24744 (2016).
\bibitem{Walzer2007} K. Walzer, B. Maennig, M. Pfeiffer, and K. Leo, Chem. Rev. $\bf{107}$, 1233 (2007).
\bibitem{Chuang2014} S. Chuang, C. Battaglia, A. Azcatl, S. McDonnell, J. S. Kang, X. Yin, M. Tosun, R. Kapadia, H. Fang, R. M. Wallace, and A. Javey, Nano Lett. $\bf{14}$, 1337 (2014).
\bibitem{Heimel2008} G. Heimel, L. Romaner, E. Zojer, and J. -L. Bredas, Acc. Chem. Res. $\bf{41}$, 721 (2008).
\bibitem{Bloom2003} C. J. Bloom, C. M. Elliott, P. G. Schroeder, C. B. France, and B. A. Parkinson, J. Phys. Chem. B $\bf{107}$, 2933 (2003).
\bibitem{Matz2013} D. L. Matz, E. L. Ratcliff, J. Meyer, A. Kahn, and J. E. Pemberton, ACS Appl. Mater. Interfaces $\bf{5}$, 6001 (2013).
\bibitem{Zhou2012} Y. Zhou, C. Fuentes-Hernandez1, J. Shim, J. Meyer, A. J. Giordano, H. Li, P. Winget, T. Papadopoulos, H. Cheun, J. Kim, M. Fenoll,
A. Dindar, W. Haske, E. Najafabadi, T. M. Khan, H. Sojoudi, S. Barlow, S. Graham, J. -L. Br{\'e}as, S. R. Marder, A. Kahn, B. Kippelen, Science $\bf{336}$, 327 (2012).
\bibitem{Jorgensen2008} M. J{\o}rgensen, K. Norrman, F. C. Krebs, Sol. Energy Mater. Sol. Cells $\bf{92}$, 686 (2008).
\bibitem{Braun2009} S. Braun, W. R. Salaneck, and M. Fahlman, Adv. Mater. $\bf{21}$, 1450 (2009).
\bibitem{Khan2014}T. M. Khan, Y. Zhou, A. Dindar, J. W. Shim, C. FuentesHernandez, and B. Kippelen, ACS Appl. Mater. Interfaces $\bf{6}$, 6202 (2014).
\bibitem{Ratcliff2011}  E. L. Ratcliff, B. Zacher, and N. R. Armstrong, J. Phys. Lett. $\bf{2}$, 1337 (2011).
\bibitem{Koch2007} N. Koch, ChemPhysChem $\bf{8}$, 1438 (2007).
\bibitem{Kano2009} M. Kano, T. Minari, and K. Tsukagoshi, Appl. Phys. Lett. $\bf{94}$, 143304 (2009).
\bibitem{Chen2008} M.-H. Chen and C.-I Wu, J. Appl. Phys. $\bf{104}$, 113713 (2008).
\bibitem{deBoer2005} B. deBoer, A. Hadipour, M. M. Mandoc, T. van Woudenbergh, P. W. M. Blom, Adv. Mater. $\bf{17}$, 621 (2005).
\bibitem{Deckman2015} I. Deckman, S. Obuchovsky, M. Moshonov, and G. L. Frey, Langmuir $\bf{31}$, 6721 (2015).
\bibitem{Shamieh2016} B. Shamieh, S. Obuchovsky, and G. L. Frey, J. Mater. Chem. C $\bf{4}$, 1821 (2016).
\bibitem{Vinokur2017} J. Vinokur, S. Obuchovsky, l. Deckman, L. Shoham, T. Mates, M. L. Chabinyc, G. L. Frey, Appl. Mater. Interfaces $\bf{9}$ , 29889 (2017).
\bibitem{Nouzman2017} L. Nouzman and G. L. Frey, J. Mater. Chem. C $\bf{5}$, 12744 (2017).
\bibitem{Sarkar2020} T. Sarkar, B. Shamieh, R. Verbeek, A. J. Kronemeijer, G. H. Gelinck, and G. L. Frey, Adv. Funct. Mater. $\bf{30}$, 1805617 (2020).
\bibitem{Shamieh2018} B. Shamieh, A. S. Anselmo, U. Vogel, E. Lariou, S. C. Hayes, N. Koch, and G. L. Frey, J. Mater. Chem. C $\bf{6}$, 8060 (2018).
\bibitem{Heimel2010} G. Heimel, F. Rissner, and E. Zojer, Adv. Mater. $\bf{22}$, 2494 (2010).
\bibitem{Love2005} J. C. Love, L. A. Estroff, J. K. Kriebel, R. G. Nuzzo, and G. M. Whitesides, Chem. Rev. $\bf{105}$, 1103 (2005).
\bibitem{Tao2007} F. Tao and S. L. Bernasek, Chem. Rev. $\bf{107}$, 1408 (2007).
\bibitem{Lee2015} H. J. Lee, A. C. Jamison, and T. R. Lee, Acc. Chem. Res. $\bf{48}$, 3007 (2015).
\bibitem{Vericat2001} C. Vericat, G. Andreasen, M. E. Vela, H. Martin, and R. C. Salvarezza, J. Chem. Phys. $\bf{115}$, 6672 (2001).
\bibitem{Vericat2005} C. Vericat, M. E. Vela, and R. C. Salvarezza, Phys. Chem. Chem. Phys. $\bf{7}$, 3258 (2005).
\bibitem{Azzam2003} {\black W. Azzam, C. Fuxen, A. Birkner, H.-T. Rong, M. Buck, and C. W{\"o}ll, Langmuir $\bf{19}$, 4958 (2003).}
\bibitem{Gulans2014} A. Gulans, S. Kontur, C. Meisenbichler, D. Nabok, P. Pavone, S. Rigamonti, S. Sagmeister, U. Werner, C. Draxl, J. Phys.: Condens. Matter. $\bf{26}$, 363202 (2014).
\bibitem{Perdew1996} J. P. Perdew, K. Burke, and M. Ernzerhof, Phys. Rev. Lett. $\bf{77}$, 3865 (1996).
\bibitem{Grimme2006} S. Grimme, J. Comput. Chem. $\bf{27}$, 1787 (2006).
\bibitem{Ambrosetti2014} A. Ambrosetti, A. M. Reilly, R. A. DiStasio Jr., and A. Tkatchenko, J. Chem. Phys. $\bf{140}$, 18A508 (2014).
\bibitem{Bucko2016} T. Bu{\v{c}}ko, S. Leb{\`e}gue, T. Gould, and J. G. {\'A}ngy{\'a}n,  J. Phys.: Condens. Matter. $\bf{28}$, 045201 (2016).
\bibitem{Tkatchenko2009} A. Tkatchenko and M. Scheffler, Phys. Rev. Lett. $\bf{102}$, 073005 (2009).
\bibitem{Nara2004} J. Nara, S. Higai, Y. Morikawa, and  T. Ohno, J. Chem. Phys. $\bf{120}$, 6705 (2004).
\bibitem{Tonigold2013} K. Forster-Tonigold, X. Stammer, C. W{\"o}ll, and A. Gro{\ss}, Phys. Rev. Lett. $\bf{111}$, 086102 (2013).
\bibitem{Tonigold2015} K. Forster-Tonigold and A. Gro{\ss}, Surface Science $\bf{640}$, 18 (2015).
\bibitem{Wu2009}  K. Y. Wu, S. Y. Yu, and Y. T. Tao, Lanmuir $\bf{25}$, 6232 (2009).
\bibitem{Tao2011}  Y.T. Tao, K.Y. Wu, K. H. Huang, and T. P. Perng, Organic Electronics $\bf{12}$, 602 (2011).
\bibitem{Duhm2008} S. Duhm, A. Gerlach, I. Salzmann, B. Br{\"o}ker, R.L. Johnson, F. Schreiber, and N. Koch, Org. Electron. $\bf{9}$, 111 (2008).
\bibitem{Miller2009} N. E. Singh-Miller and N. Marzari, Phys. Rev. B $\bf{80}$, 235407 (2009).
\bibitem{Patra2017} A. Patra, J. E. Bates, J. Sun, and J. P. Perdew, PNAS $\bf{114}$, E9188 (2017).
\bibitem{Li2010} H. Li, P. Paramonov, and J. L. Bredas, J. Mater. Chem. $\bf{20}$, 2630 (2010).
\bibitem{Cornil2014} D. Cornil, T. Van Regemorter, D. Beljonne, and J. Cornil, Phys. Chem. Chem. Phys. $\bf{16}$, 20887 (2014).
\bibitem{Wang2019} Q. Wang, V. Diez-Cabanes, S. Dell'Elce, A. Liscio, B. Kobin, H. Li, J. Br{\`e}das, S. Hecht, V. Palermo, E. J. W. List-Kratochvil, J. Cornil, N. Koch, and G. Ligorio,
ACS Appl. Nano Mater. $\bf{2}$, 1102 (2019).
\bibitem{Ulman1996} A. Ulman, Chem. Rev. $\bf{96}$, 1533 (1996).
\bibitem{Sony2007} P. Sony, P. Puschnig, D. Nabok, and C. Ambrosch-Draxl, Phys. Rev. Lett. $\bf{99}$, 176401 (2007).
\bibitem{Romaner2009} L. Romaner, D. Nabok, P. Puschnig, E. Zojer, and C. Ambrosch-Draxl, New J. Phys. $\bf{11}$, 053010 (2009).
\bibitem{Tkatchenko2010} A. Tkatchenko, L. Romaner, O. T. Hofmann, E. Zojer, C. Ambrosch-Draxl, and M. Scheffler, MRS Bulletin $\bf{35}$, 435 (2010).
\bibitem{Nabok2008} D. Nabok, P. Puschnig and C. Ambrosch-Draxl, Phys. Rev. B $\bf{77}$, 245316 (2008).
\end{thebibliography}

\end{document}